\documentclass[aps, prl, reprint, superscriptaddress]{revtex4-2}

\usepackage[utf8]{inputenc}
\usepackage[english]{babel}
\usepackage[colorlinks, urlcolor=blue, citecolor=blue]{hyperref}
\usepackage{graphicx}
\usepackage{amssymb, amsmath}
\usepackage{siunitx}
\usepackage{verbatim}

\DeclareMathOperator{\csch}{csch}

\newcommand{\avg}[1]{\ensuremath{\langle #1\rangle}}
\newcommand{\expct}[3]{\ensuremath{\langle #1|#2|#3 \rangle}}
\newcommand{\sprod}[2]{\ensuremath{\langle #1|#2\rangle}}
\newcommand{\ket}[1]{\ensuremath{|#1\rangle}}
\newcommand{\bra}[1]{\ensuremath{\langle #1|}}

\begin{document}

\title{Swap-test interferometry with biased ancilla noise}

\author{Ond\v{r}ej \v{C}ernot\'ik}
\email{ondrej.cernotik@upol.cz}
\affiliation{Department of Optics, Palack\'y University, 17. listopadu 1192/12, 77146 Olomouc, Czechia}

\author{Iivari Pietik\"{a}inen}
\affiliation{Department of Optics, Palack\'y University, 17. listopadu 1192/12, 77146 Olomouc, Czechia}

\author{Shruti Puri}
\affiliation{Yale Quantum Institute, PO Box 208 334, 17 Hillhouse Ave, New Haven, CT 06520-8263 USA}
\affiliation{Department of Applied Physics, Yale University}

\author{S. M. Girvin}
\affiliation{Yale Quantum Institute, PO Box 208 334, 17 Hillhouse Ave, New Haven, CT 06520-8263 USA}
\affiliation{Department of Physics, Yale University}

\author{Radim Filip}
\affiliation{Department of Optics, Palack\'y University, 17. listopadu 1192/12, 77146 Olomouc, Czechia}

\date{\today}

\begin{abstract}
	The Mach--Zehnder interferometer is a powerful device for detecting small phase shifts between two light beams.
	Simple input states---such as coherent states or single photons---can reach the standard quantum limit of phase estimation while more complicated states can be used to reach Heisenberg scaling;
	the latter, however, require complex states at the input of the interferometer which are difficult to prepare.
	The quest for highly sensitive phase estimation therefore calls for interferometers with nonlinear devices which would make the preparation of these complex states more efficient.
	Here, we show that the Heisenberg scaling can be recovered with simple input states (including Fock and coherent states) when the linear mirrors in the interferometer are replaced with controlled-swap gates and measurements on ancilla qubits.
	These swap tests project the input Fock and coherent states onto NOON and entangled coherent states, respectively, leading to improved sensitivity to small phase shifts in one of the interferometer arms.
	We perform detailed analysis of ancilla errors, showing that biasing the ancilla towards phase flips offers a great advantage, and perform thorough numerical simulations of a possible implementation in circuit quantum electrodynamics.
	Our results thus present a viable approach to phase estimation approaching Heisenberg-limited sensitivity.
\end{abstract}

\maketitle

\begin{figure}
	\centering
	\includegraphics[width=\linewidth]{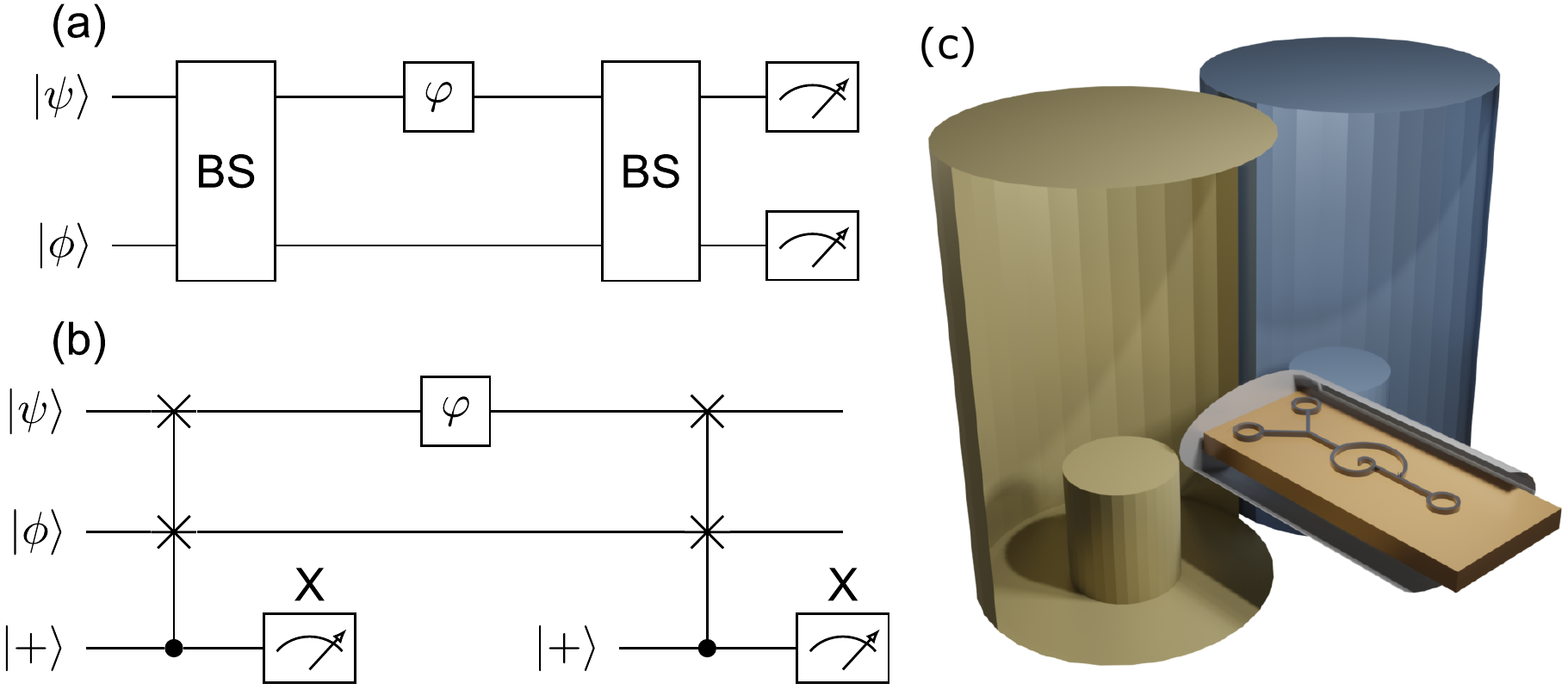}
	\caption{\label{fig:Imetry}Swap-test interferometry.
	(a) Mach--Zehnder interferometer for estimating an unknown phase $\varphi$.
	Two electromagnetic modes in quantum states $\ket{\psi},\ket{\phi}$ are superimposed on linear beam splitters (BS) sandwiching a phase shift $\varphi$ on one of the fields.
	Subsequent measurement of the output fields can be used to estimate this phase shift.
	(b) Interferometry based on swap tests.
	Instead of linear beam splitters, controlled-swap gates with an ancilla qubit (initially in the state $\ket{+} = (\ket{0}+\ket{1})/\sqrt{2}$) are used;
	the detection of the fields is replaced by measurement of the ancilla in the $X$ basis.
	(c) Depiction of a possible experimental implementation in circuit QED.
	Two three-dimensional microwave cavity modes interact with a SNAIL circuit.
	The SNAIL is used to implement a Kerr-cat qubit which controls the swapping of the two fields.
	Using a Kerr-cat qubit instead of a transmon leads to biasing of the ancilla noise which helps to ensure high phase sensitivity at the Heisenberg limit.
	}
\end{figure}

Interferometry encompasses a broad range of devices and techniques that use the wave nature of quantum systems to estimate small phase shifts~\cite{DemkowiczDobrzanski2015,Degen2017}.
While various interferometer topologies and architectures exist, their operational principle remains the same---a probe beam is subject to a phase shift which is estimated by analysing interference with a reference beam.
Among the different interferometer designs, the Mach--Zehnder interferometer (see Fig.~\ref{fig:Imetry}(a)) is often used not only for highly accurate estimation of unknown phases~\cite{Giovannetti2011,Polino2020} but has also found use in quantum computing applications~\cite{OBrien2003,Crespi2013,Micuda2013,Peruzzo2014}.
Due to the linearity of the Mach--Zehnder interferometer, sensitivity of phase estimation is limited by the standard quantum limit for simple input states (such as single photons and coherent states), which scales as $1/\sqrt{n}$, where $n$ is the number of photons used~\cite{Pezze2007}.
Improvements beyond the standard quantum limit (going all the way to Heisenberg scaling, $1/n$~\cite{Pezze2008,Giovannetti2004}) are possible with more complex states, such as NOON states, which are entangled states of the $n$-photon Fock state with the vacuum, $(\ket{n}\ket{0}+\ket{0}\ket{n})/\sqrt{2}$~\cite{Walther2004}, and entangled coherent states, $(\ket{\alpha_1}\ket{\alpha_2}+\ket{\alpha_2}\ket{\alpha_1})/\sqrt{N_+}$, where $\alpha_{1,2}$ are two coherent-state amplitudes and $N_+$ is a normalisation constant~\cite{Joo2011}.

Preparation and measurement of these complex quantum states is, however, far from trivial.
While two-photon NOON states can be prepared using Hong--Ou--Mandel interference---in which putting one photon in each of the two input modes results in photon bunching and both photons leaving through the same output mode---creation of NOON states with higher photon numbers requires complex operations with efficient photodetection and feedforward~\cite{Cable2007} or building the target state one excitation at a time~\cite{Zhang2018}.
In addition, efficient phase estimation with NOON states requires photon-number (or phonon-number) resolving detectors which are not available for standard optical or trapped-ion systems.
Spatial~\cite{Hlousek2019} or time multiplexing~\cite{Achilles2006} is required for optical photons while trapped-ion systems use complex control schemes~\cite{Ding2017,Wolf2019,Ohira2019};
in all these approaches, the amount of resources needed scales unfavourably with the size of the NOON state.
Photon counting can be achieved in circuit QED using dispersive interaction of a microwave mode with an ancilla qubit but this approach also requires complicated control schemes~\cite{Wang2020,Curtis2021}.
Phase estimation would therefore profit from alternative approaches to creating and measuring these highly sensitive quantum states.

The limits posed by standard linear Mach--Zehnder interferometers can be overcome with the help of two-mode squeezers replacing beam splitters~\cite{Yurke1986,Tse2019,Backes2021} or nonlinear interferometry.
In such a scenario, nonlinearity can be introduced in one (or both) of the interferometer arms~\cite{Gerry2002}, instead of the linear beam splitters that mix the two modes~\cite{Leibfried2002,Gross2010,Ou2012,Hudelist2014}, or in measurement of the output states~\cite{Sewell2014}.
While such strategies provide advantages over linear interferometry, these techniques often rely on strong nonlinearities which are difficult to engineer or are applicable only to a specific type of quantum system.
In contrast, swap gates provide highly nonlinear interactions and, although challenging, are available in a broad range of experimental platforms~\cite{Patel2016,Starek2018,Gao2019,Zhang2019a,Gan2020} and are thus ideally suited for nonlinear interferometry.
The controlled-swap gate is a particularly attractive gate as it is an essential ingredient of swap tests which are useful for measuring properties of quantum states without full tomography~\cite{Ekert2002}, in particular state overlap and purity~\cite{Filip2002,Nguyen2021ar}, and other verification tasks~\cite{Carrasco2021}.
In addition, swap tests conditionally project the input states onto their symmetric or antisymmetric component, allowing the preparation of NOON states $(\ket{n}\ket{0}\pm\ket{0}\ket{n})/\sqrt{2}$ from input Fock states $\ket{n},\ket{0}$ and of entangled coherent states $(\ket{\alpha_1}\ket{\alpha_2}\pm\ket{\alpha_2}\ket{\alpha_1})/\sqrt{N_\pm}$ from coherent states $\ket{\alpha_{1,2}}$.
Finally, the same principles allow swap tests to be used for high-fidelity measurements of these complex quantum states~\cite{Filip2002}.

In this article, we propose a nonlinear extension of the Mach--Zehnder interferometer in which conventional linear beam splitters have been replaced with swap tests (Fig.~\ref{fig:Imetry}(b)).
The first swap test is used to conditionally prepare a state that probes an unknown phase shift $\varphi$ by projecting the input states $\ket{\psi},\ket{\phi}$ onto their symmetric or antisymmetric component.
This setup prepares the entangled NOON state from the Fock state $\ket{n}$ and the vacuum or an entangled coherent state from two coherent states;
we discuss the difference between these scenarios caused by the finite overlap of the two coherent states $\sprod{\alpha_1}{\alpha_2}\neq 0$.
The second swap test is then used to estimate a phase shift on one of the modes from the probability that an antisymmetric state will be projected onto the symmetric subspace or vice versa.
In trapped-ion and superconducting systems, where the controlled-swap gates are readily available~\cite{Gao2019,Gan2020,Nguyen2021ar}, interferometry with NOON states becomes much more straightforward:
a NOON state can be both prepared and measured in a single step irrespective of its size.
The only resource needed is then an input Fock state $\ket{n}$, for which efficient preparation methods exist in both circuit QED~\cite{Heeres2015,Heeres2017,Eickbusch2021X} and trapped-ion systems~\cite{McCormick2019,Podhora2021X}.

To provide a complete picture of swap-test interferometry in realistic conditions, we evaluate errors of the ancilla qubits, namely phase and bit flips, and show that the two types of error play a fundamentally different role.
Phase flips result in incorrect assignment of the measurement results to projections of the field states onto the symmetric and antisymmetric subspace, reducing the overall interference contrast;
owing to the nondemolition nature of the swap test, repeated swap tests with the same ancilla and cavity fields can be used to correct for these errors.
Ancilla bit flips during the controlled-swap gate, on the other hand, lead to over- and underrotation of the two-mode state during the swap.
Even though the measurement still projects the modes onto their symmetric and antisymmetric components, the generated states are different from the ideal NOON and entangled coherent states.
These errors---which cannot be detected with repeated swap tests---therefore limit the estimation sensitivity and prevent us from reaching the Heisenberg limit.

To overcome the limitation posed by ancilla bit flips, we discuss a possible implementation in circuit quantum electrodynamics (Fig.~\ref{fig:Imetry}(c)).
Here, the swap operation between the two cavity modes is controlled by a Kerr-cat qubit which exhibits strong noise bias~\cite{Puri2017,Grimm2020}.
In this type of qubit, photon loss (the dominant decoherence mechanism) introduces phase flips while bit flips are exponentially suppressed~\cite{Puri2019,Puri2020}.
Suitable driving can then be used to engineer a beam-splitter interaction between the two cavity fields controlled by this ancilla cat qubit~\cite{Pietikainen2020}.
We perform detailed numerical simulations of the whole protocol to (i) analyse the overlap witness detecting nonclassical correlations between the two modes in a realistic setting with losses and noise and (ii) confirm that the standard quantum limit can be surpassed in these realistic devices and the Heisenberg limit is attainable.
Our work thus presents an attractive target for experiments in quantum enhanced phase estimation with available technology.

\section{Results}

\subsection{Swap-test interferometry}

A swap test can be implemented using a controlled-swap gate between two fields controlled by an ancilla qubit.
The qubit is initially prepared in the state $\ket{+} = (\ket{0}+\ket{1})/\sqrt{2}$ and its state is measured after the gate in the $X$ basis.
This process projects the input state of the fields onto its symmetric (for measurement outcome $\ket{+}$) or antisymmetric [for $\ket{-} = (\ket{0}-\ket{1})/\sqrt{2}$] subspace, defined by the projectors $\Pi_+ = (I+S)/\sqrt{2}$ (symmetric) and $\Pi_- = (I-S)/\sqrt{2}$ (antisymmetric), where $I$ is the identity and $S\ket{\psi}\ket{\phi} = \ket{\phi}\ket{\psi}$ is the swap operator~\cite{Filip2002}.
For two general, orthogonal quantum states $\ket{\psi}$, $\ket{\phi}$ (satisfying $\sprod{\psi}{\phi} = 0$), the swap test conditionally prepares one of the two Bell states
\begin{equation}
	\ket{\Psi_\pm} = \frac{1}{\sqrt{2}}(\ket{\psi}\ket{\phi}\pm\ket{\phi}\ket{\psi})
\end{equation}
with probability $p_\pm = \frac{1}{2}$.
If, on the other hand, one of the Bell states $\ket{\Psi_\pm}$ is at the input of the swap test, only one measurement outcome is possible---for the symmetric state $\ket{\Psi_+}$, the ancilla is always in the state $\ket{+}$ whereas for the antisymmetric state $\ket{\Psi_-}$ it is always in the state $\ket{-}$.

The proposed swap-test interferometry uses two such swap tests sandwiching a phase shift on one of the two modes as shown in Fig.~\ref{fig:Imetry}(b).
With the fields starting in two Fock states $\ket{n}$, $\ket{m}$, $n\neq m$, the first swap test conditionally prepares one of the Bell states
\begin{equation}
	\ket{\Psi_\pm} = \frac{1}{\sqrt{2}}(\ket{n}\ket{m} \pm \ket{m}\ket{n}).
\end{equation}
Focusing, for the moment, on the antisymmetric state $\ket{\Psi_-}$, we obtain the following state after the phase shift on the first mode:
\begin{equation}
	\ket{\Psi_-(\varphi)} = \frac{1}{\sqrt{2}}(e^{-in\varphi}\ket{n}\ket{m} - e^{-im\varphi}\ket{m}\ket{n}).
\end{equation}
The second swap test is then used to determine the difference of the probabilities $p_\pm$ of detecting the ancilla in the states $\ket{\pm}$,
\begin{equation}
	\Delta(\varphi) = p_+ - p_- = -\cos[(n-m)\varphi].
\end{equation}
Since $p_- = 1$ for an antisymmetric state, the quantity $\Delta(\varphi)$ serves as a witness of singlet-like entanglement in the fields~\cite{Filip2002}.

With the choice $m = 0$, the first swap test conditionally prepares the NOON state $\ket{\Psi_-} = (\ket{n}\ket{0}-\ket{0}\ket{n})/\sqrt{2}$ and the witness becomes
\begin{equation}
	\Delta(\varphi) = -\cos(n\varphi).
\end{equation}
The swap-test interferometer can thus be used for estimating the phase $\varphi$ with Heisenberg scaling with a simple input state $\ket{n}\ket{0}$.
A Mach--Zehnder interferometer, on the other hand, would need a complicated superposition at the input for $n>2$;
its precise form can be found by propagating the desired NOON state backwards through the balanced linear beam splitter (see Fig.~\ref{fig:Imetry}(a)).
This advantage is enabled by the nonlinear transformation in the swap test which, however, can be efficiently implemented in circuit QED~\cite{Pietikainen2020} or with trapped ions~\cite{Gan2020}.

An important issue that could easily quell any advantage that swap-test interferometry can provide over conventional Mach--Zehnder interferometers are errors of the ancilla qubits.
First, phase flips result in systematic errors in assigning the measurement outcome and associated projection onto the symmetric or antisymmetric subspace.
Since the gate is transparent to phase-flip errors (which are described by the Pauli $Z$ operator; the gate is given by the unitary $\frac{1}{2}(I-Z)\otimes S$ and therefore commutes with the error), we can consider only phase-flip errors that occur after the gate and before the measurement.
It can be shown (see Methods) that the incorrect assignment of a measurement outcome to a projection reduces the interference visibility by a factor $1-2p$, where $p$ is the probability of a phase-flip error.
In addition, given the transparency of the controlled-swap gate to this type of error, it can be accounted for by repeating the swap test on the same state and then taking the majority vote.
While sucha a scheme can improve the sensitivity in principle (provided the phase-flip probability $p<\frac{1}{2}$), other effects (such as cavity losses or limited efficiency of the qubit measurement) might limit its applicability.

Bit-flip errors, on the other hand, pose a more serious threat:
while they do not affect the protocol when they happen before or after the controlled-swap gate (since the initial state is an eigenstate of the Pauli $X$ operator and the final measurement is performed in the $X$ basis),
a bit flip during the controlled-swap gate results in imperfect swap, scrambling the output state.
The specifics of such a process depend on the precise implementation of the gate and the exact timing of the error but will generally lead to an under- or over-rotation of the swap gate, giving rise to a more general beam-splitter-like transformation of the cavity fields with modified amplitude of the transmission and reflection coefficients.
Starting with Fock-state input, a general superposition of different Fock states in the two modes will be created instead of a NOON state, making Heisenberg scaling unattainable.

\subsection{Overlap witness with general pure states}

So far, we have assumed that the two states at the input of the swap-test interferometer are orthogonal.
While this is the case for Fock states (with which the NOON states can be created and the Heisenberg scaling reached), for other important classes of states---such as coherent states---this is not the case.
Therefore, we now turn our attention to general pure states at the input with a finite overlap $\sprod{\psi}{\phi} = s\in\mathbb{C}$.
As we shall see, even coherent states allow sensitivity of phase estimation at the Heisenberg limit;
linear interferometers, on the other hand, are bounded by the standard quantum limit with coherent-state input.
This remarkable effect is enabled by the fact that the controlled-swap gate turns the coherent states into an entangled coherent state.

While the first swap test still projects the two modes onto their symmetric or antisymmetric component, the respective probability is modified due to the finite overlap between the states $\ket{\psi},\ket{\phi}$.
This results in different normalisation for the two Bell-like states,
\begin{equation}
	\ket{\Psi_\pm} = \frac{1}{\sqrt{N_\pm}}(\ket{\psi}\ket{\phi} \pm \ket{\phi}\ket{\psi})
\end{equation}
with the normalisation factors given by $N_\pm = 2(1\pm|s|^2)$.
This, in turn, results in different probabilities of preparing these states, $p_\pm = \frac{1}{2}(1\pm|s|^2)$.
The state (we again assume that the singlet-like state $\ket{\Psi_-}$ has been prepared) is then subject to the phase shift and the second swap test, after which we obtain the overlap witness (see Methods)
\begin{align}\label{eq:witness}
	\Delta(\varphi) &= p_+ - p_-  \\
	&= \frac{|s(\varphi)|^2+|s(-\varphi)|^2 -s_\psi(\varphi)s_\phi(-\varphi) -s_\psi(-\varphi)s_\phi(\varphi)}{2(1-|s|^2)}, \nonumber
\end{align}
where we introduce the notation
\begin{align}\label{eq:products}
\begin{split}
	s(\varphi) &= \sprod{\phi}{\psi(\varphi)} = \expct{\phi}{e^{-i\varphi a^\dagger a}}{\psi}, \\
	s_\chi(\varphi) &= \sprod{\chi}{\chi(\varphi)} = \expct{\chi}{e^{-i\varphi a^\dagger a}}{\chi},
\end{split}
\end{align}
where $a$ is the annihilation operator of the first mode and $\chi = \psi,\phi$.

The more generic interference pattern possible with the overlap witness~\eqref{eq:witness} gives rise to a plethora of possible phase-estimation scenarios with high sensitivity for simple input states.
As a concrete example, we now consider two coherent states $\ket{\alpha_{1,2}}$ with the scalar product
\begin{equation}
	s = \sprod{\alpha_1}{\alpha_2} = \exp\left(-\frac{1}{2}|\alpha_1|^2-\frac{1}{2}|\alpha_2|^2 + \alpha_1^\ast\alpha_2\right).
\end{equation}
The expression for the overlap witness with two general coherent states at the input is cumbersome and offers little insight so we focus on two specific regimes:
two states with equal but opposite amplitudes, $\alpha_1 = -\alpha_2 = \alpha$, and a coherent state with a vacuum, $\alpha_1 = \alpha$, $\alpha_2 = 0$.
In both cases, we take $\alpha\in\mathbb{R}$ without loss of generality.

\begin{figure}
	\centering
	\includegraphics[width=\linewidth]{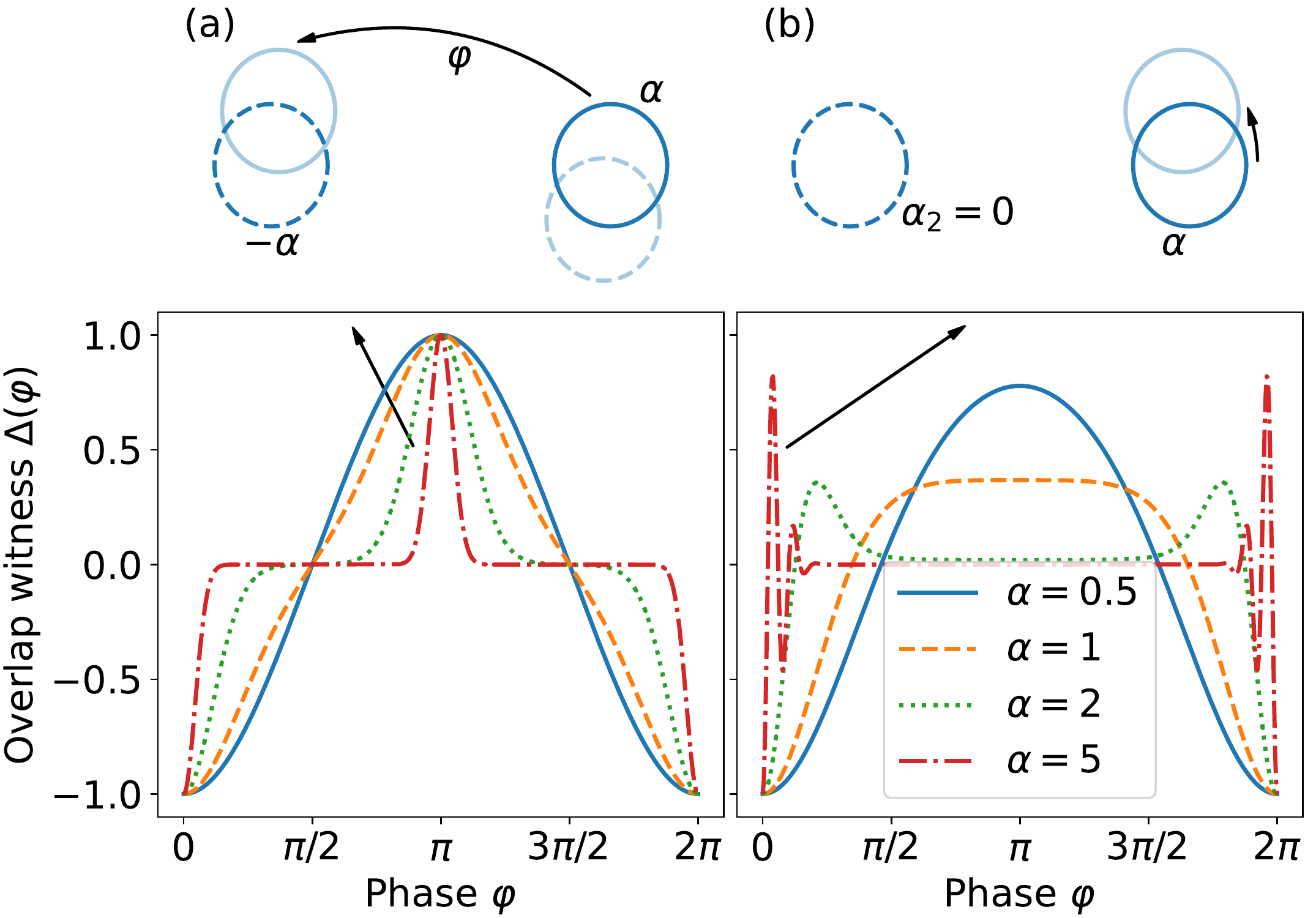}
	\caption{\label{fig:coherent}
		Schematic depiction of state overlap (top) and overlap witness $\Delta(\varphi)$ (bottom) with coherent states.
		(a) Overlap witness for coherent states with amplitudes $\alpha_1 = -\alpha_2 = \alpha$.
		For large amplitudes $\alpha\gg 1$, significant overlap between the original and phase-shifted states occurs only for $\varphi \sim k\pi$, with $k\in\mathbb{Z}$, giving rise to a plateau in between.
		(b) Overlap witness for a coherent state $\ket{\alpha}$ and the vacuum.
		Only the coherent state $\ket{\alpha}$ rotates under a phase shift;
		the interference pattern between the superposed states $\ket{0},\ket{\alpha}$ (not shown) gives rise to fast oscillations of the overlap witness $\Delta(\varphi)$.}
\end{figure}

For two opposite-amplitude coherent states, a straightforward calculation gives the overlap witness
\begin{equation}\label{eq:catWitness}
	\Delta(\varphi) = -\frac{\sinh(2\alpha^2\cos\varphi)}{\sinh(2\alpha^2)},
\end{equation}
whose interference pattern is shown in Fig.~\ref{fig:coherent}(a).
The overlap witness satisfies $\Delta(\varphi) = -1$ for $\varphi = 2k\pi$ and $\Delta(\varphi) = 1$ for $\varphi = (2k+1)\pi$, where $k\in\mathbb{Z}$.
The speed with which the overlap witness drops to zero depends on the amplitude $\alpha$ as can be seen from the following argument:
The witness effectively measures the overlap between the initial coherent states $\ket{\pm\alpha}$ and their phase-shifted variant, $\ket{\pm\alpha e^{-i\varphi}}$.
As the amplitude $\alpha$ increases, the range of phases for which these states significantly overlap decreases [see top of Fig.~\ref{fig:coherent}(a)].
When the original and phase-shifted states do not overlap, projections on the symmetric and antisymmetric subspaces are equally likely and we have $\Delta(\varphi) = 0$.

For a coherent state and the vacuum, the overlap witness takes the form
\begin{equation}\label{eq:Delta_a0}
	\Delta(\varphi) = -\frac{1 - \exp(\alpha^2\cos\varphi)\cos(\alpha^2\sin\varphi)}{1-\exp(\alpha^2)},
\end{equation}
which shows fast oscillatory pattern for large amplitudes $\alpha$ around $\varphi = 2k\pi$ (see Fig.~\ref{fig:coherent}(b)).
The main difference from the previous case responsible for this behaviour is the different rotation axis.
With $\alpha_2 = -\alpha_1$, phase rotation corresponds to a rotation of the whole state around its centre,
whereas for $\alpha_2 = 0$, the rotation axis is located at the centre of the coherent state $\ket{\alpha_2} = \ket{0}$ in phase space.
For large amplitudes, there are now two main contributions to the overlap between the initial and phase-shifted states.
The first is, again, the overlap between the coherent state $\ket{\alpha}$ and its phase-shifted variant $\ket{\alpha e^{-i\varphi}}$ which gives an envelope of the overlap witness.
The oscillations under this envelope are caused by the rapidly changing overlap in the interference pattern between the states $\ket{\alpha_1} = \ket{\alpha}$, $\ket{\alpha_2} = \ket{0}$.
For large coherent amplitudes, even a small phase shift easily turns the troughs in this interference region into ridges and vice versa, resulting in approximately orthogonal states with $\Delta(\varphi)\to 1$ (red dot-dashed line).

For both Fock and coherent states, we assumed that the antisymmetric, singlet-like state $\ket{\Psi_-} = (\ket{\psi}\ket{\phi}-\ket{\phi}\ket{\psi})/\sqrt{N_-}$ was prepared in the first swap test but the analysis can be repeated for the symmetric state $\ket{\Psi_+} = (\ket{\psi}\ket{\phi}+\ket{\phi}\ket{\psi})/\sqrt{N_+}$.
Irrespective of the input states, these symmetric and antisymmetric superpositions differ only by a relative phase that can be taken into account when evaluating the overlap witness $\Delta$.
As long as one keeps a record of both measurement outcomes, and evaluates the overlap witness for symmetric and antisymmetric correlations separately, the symmetric and anti-symmetric states can be jointly used to estimate the unknown phase $\varphi$.

\subsection{Fisher information}

To gain more insight into the sensitivity of the swap-test interferometer, we now evaluate the Fisher information for the different types of input states discussed above.
We will calculate the quantum Fisher information of the state of the cavity modes after the phase shift, so either
\begin{equation}
	\ket{\Psi_-(\varphi)} = \frac{1}{\sqrt{2}}(e^{-in\varphi}\ket{n}\ket{0}-\ket{0}\ket{n})
\end{equation}
for NOON states, or
\begin{equation}
	\ket{\Psi_-(\varphi)} = \frac{1}{\sqrt{N_-}}(\ket{\alpha_1 e^{-i\varphi}}\ket{\alpha_2}-\ket{\alpha_2 e^{-i\varphi}}\ket{\alpha_1})
\end{equation}
for entangled coherent states, to obtain the ultimate sensitivity limit.
We also evaluate the classical Fisher information contained in the correlations between the two ancilla measurements to see how close to this limit the swap-test interferometer is.

For NOON states, a straightforward calculation (see Methods) gives
\begin{equation}
	F_Q^{\rm NOON} = F_C^{\rm NOON} = n^2,
\end{equation}
where $F_Q^{\rm NOON}$ ($F_C^{\rm NOON}$) denotes the quantum (classical) Fisher information.
The quantum Fisher information gives the ultimate sensitivity limit for NOON states via the quantum Cram\'{e}r--Rao bound~\cite{Haase2018,Polino2020}, showing the expected Heisenberg scaling of $1/n^2$.
The fact that the classical and quantum Fisher information are equal shows that swap test presents the optimal measurement strategy for NOON states~\cite{Haase2018,Polino2020}.

\begin{figure}
	\centering
	\includegraphics[width=\linewidth]{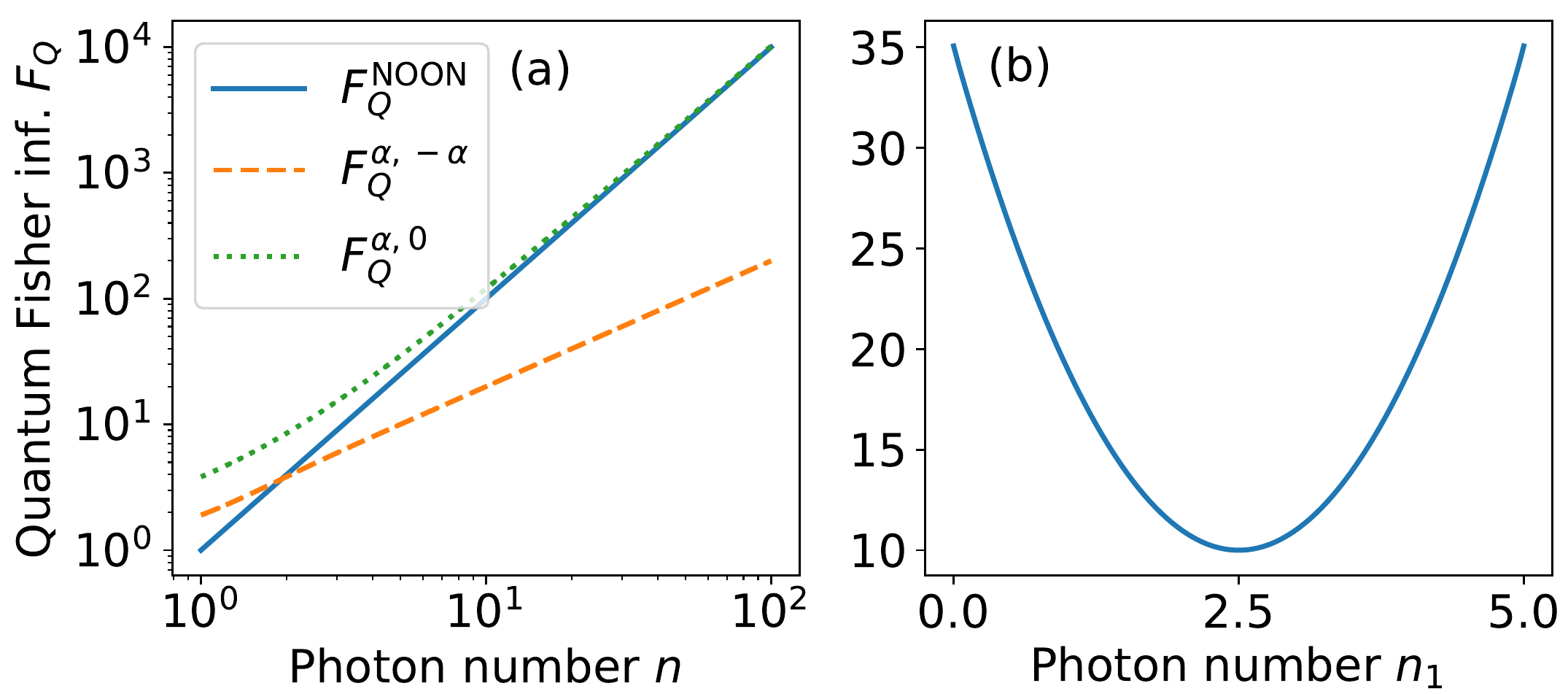}
	\caption{\label{fig:QFI} Quantum Fisher information in swap-test interferometry.
	(a) Scaling of the quantum Fisher information with the total photon number at the input for NOON states ($F_Q^{\rm NOON}$, solid blue line), entangled coherent states with $\alpha_1 = -\alpha_2$ ($F_Q^{\alpha,-\alpha}$, dashed orange line), and entangled coherent states with $\alpha_2 = 0$ ($F_Q^{\alpha,0}$, dotted green line). The total photon number for entangled coherent states is $\alpha_1^2+\alpha_2^2$.
	(b) Quantum Fisher information for entangled coherent states with the total photon number $n = 5$ for different partitions of the total energy between the two modes. We assume two real coherent-state amplitudes $\alpha_1 = \sqrt{n_1}$, $\alpha_2 = -\sqrt{n-n_1}$.}
\end{figure}

For entangled coherent states, the general formula for the quantum Fisher information $F_Q^{\alpha_1,\alpha_2}$ is more involved.
For the two special cases analysed in Fig.~\ref{fig:coherent}, we obtain (see Methods)
\begin{subequations}\label{eq:QFIecs}
\begin{align}
	F_Q^{\alpha,-\alpha} &= 2\alpha^2\csch^2(2\alpha^2)[\sinh(4\alpha^2)-2\alpha^2],\\
	F_Q^{\alpha,0} &= \frac{\alpha^2 e^{\alpha^2}}{(1-e^{\alpha^2})^2}[e^{\alpha^2}(2+\alpha^2)-2(1+\alpha^2)].
\end{align}
\end{subequations}
The quantum Fisher information for these different types of probe states is plotted in Fig.~\ref{fig:QFI}.
In panel (a), we plot the quantum Fisher information for the three sensing scenarios (i.e., NOON states and entangled coherent states with $\alpha_2 = -\alpha_1$ or $\alpha_2 = 0$) against the photon number at the input of the interferometer.
Only two of the analysed situations can give rise to Heisenberg scaling;
using two coherent states with equal but opposite amplitudes becomes $F_C^{\alpha,-\alpha}\simeq 2n$ in the limit of large photon numbers, corresponding to the standard quantum limit.
Remarkably, despite this less favourable scaling, this strategy still provides a higher quantum Fisher information than NOON states for small photon numbers.
The largest values of the quantum Fisher information are obtained with $F_Q^{\alpha,0}$ which outperforms NOON states for small photon numbers and then asymptotically approaches the Heisenberg scaling $n^2$ from above.
This behaviour can be understood from the next-to-leading-order terms in the limit of large photon number which give,
\begin{equation}
	F_Q^{\alpha,0} \approx \frac{e^{2n}}{e^{2n}-2e^n}(n^2+2n) \to n^2
\end{equation}
which holds well for $n\gtrsim 3$.
In the opposite limit, $n\to 0$, we have $F_Q^{\alpha,-\alpha} = F_Q^{\alpha,0} = 1$ which is, however, accompanied by vanishing probability of preparing the singlet state in this limit.

We further investigate the phase sensitivity with entangled coherent states in Fig.~\ref{fig:QFI}(b) where we plot the quantum Fisher information $F_Q^{\alpha_1,\alpha_2}$ with fixed total energy $n = \alpha_1^2+\alpha_2^2$ ($\alpha_{1,2}\in\mathbb{R}$) as a function of the mean photon number in the first mode, $n_1 = \alpha_1^2$.
The highest quantum Fisher information (and therefore highest phase sensitivity) is achieved in the two limiting cases $n_1 = 0$, $n_1 = n$ which correspond to the whole energy being concentrated in one of the modes (with either $\alpha_1 = \sqrt{n}$, $\alpha_2 = 0$, or $\alpha_1 = 0$, $\alpha_2 = -\sqrt{n}$);
the minimum is reached for $n_1 = \frac{1}{2}n$, corresponding to $\alpha_1 = -\alpha_2$.

\begin{figure}
	\centering
	\includegraphics[width=\linewidth]{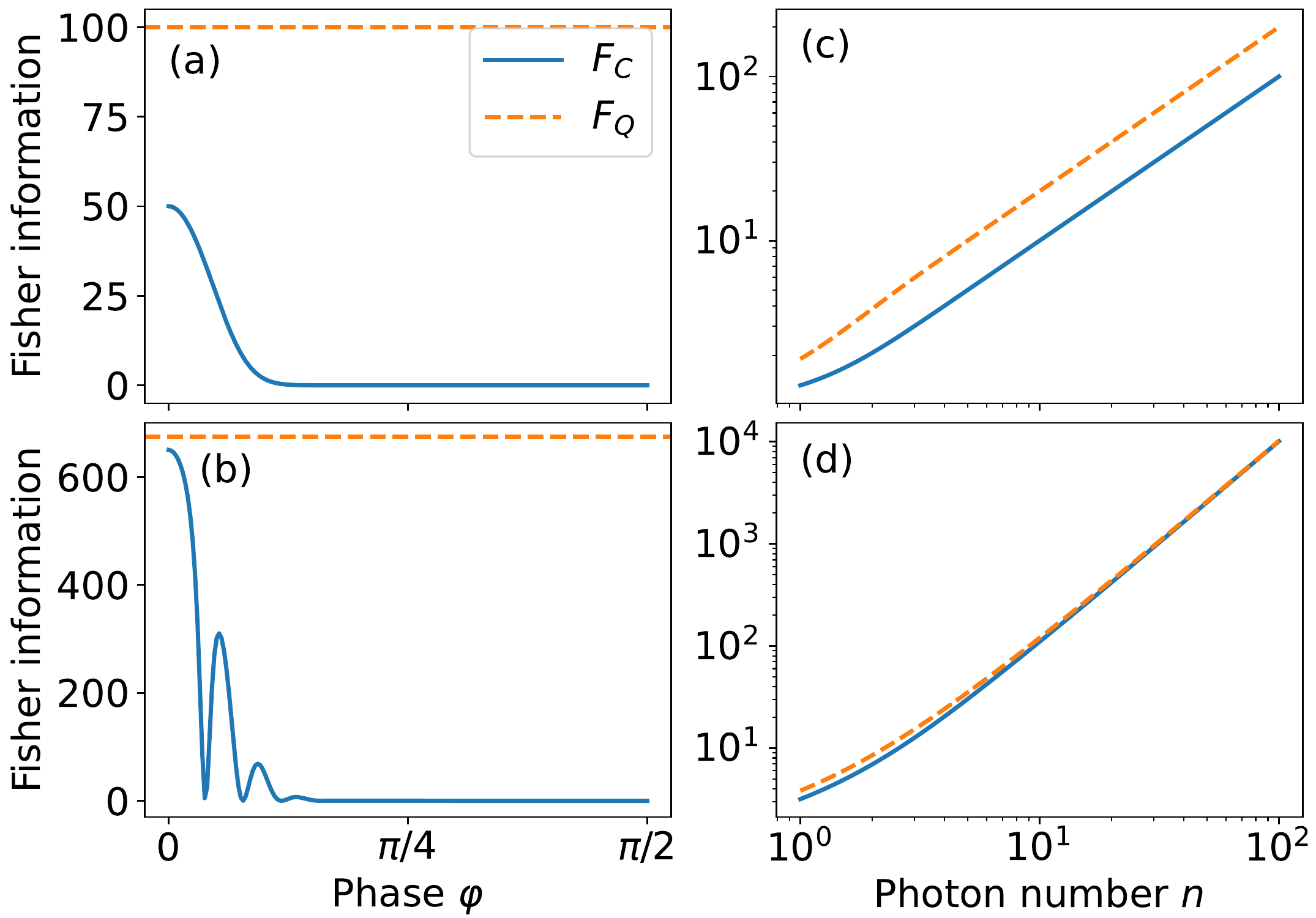}
	\caption{\label{fig:CFI} Classical Fisher information in swap-test interferometry with entangled coherent states.
	Phase dependence of the classical Fisher information for entangled coherent states with (a) $\alpha_1 = -\alpha_2 = 5$ and (b) $\alpha_1 = 5$, $\alpha_2 = 0$.
	Classical Fisher information in the limit $\varphi\to 0$ as a function of the total photon number $n = \alpha_1^2 + \alpha_2^2$ for (c) $\alpha_1 = -\alpha_2$ and (d) $\alpha_2 = 0$ ($n = \alpha_1^2$).
	In all plots, we compare the classical Fisher information (solid blue line) to the quantum Fisher information (dashed orange line).}
\end{figure}

To see how well the swap test performs when estimating the phase, we now calculate the classical Fisher information for entangled coherent states.
We plot the classical Fisher information for the cases $\alpha_1 = -\alpha_2$ and $\alpha_2 = 0$ in Fig.~\ref{fig:CFI} (see Methods for derivation and expressions).
Unlike the quantum Fisher information, the classical Fisher information is generally phase-dependent which reflects the negligible overlap between the states in a broad range of phases for large amplitudes (providing no information about the phase shift $\varphi$, cf. Fig.~\ref{fig:coherent}).
The classical Fisher information is maximal close to $\varphi = 0$;
in the limit $\varphi\to 0$, the classical Fisher information becomes
\begin{subequations}
\begin{align}
	F_C^{\alpha,-\alpha} &= 2\alpha^2\coth(2\alpha^2),\\
	F_C^{\alpha,0} &= \frac{e^{\alpha^2}(\alpha^2+\alpha^4)}{-1+e^{\alpha^2}}.
\end{align}
\end{subequations}
The former ($F_C^{\alpha,-\alpha}$) remains smaller than the corresponding quantum Fisher information;
in the large-$n$ limit, it is smaller by a factor of two, $F_C^{\alpha,-\alpha}\simeq n$ (cf. $F_Q^{\alpha,-\alpha} = 2n$).
The latter ($F_C^{\alpha,0}$) is asymptotically close to the quantum Fisher information with $F_C^{\alpha,0} \simeq F_Q^{\alpha,0} \simeq n^2$ for large photon numbers.
Unlike with NOON states, however, this high sensitivity is achievable only for a narrow range of phases in the vicinity of $\varphi = 0$.
The possibility of approaching Heisenberg scaling is, however, remarkable given the fully classical input of the swap-test interferometer.

\begin{figure}
	\centering
	\includegraphics[width=\linewidth]{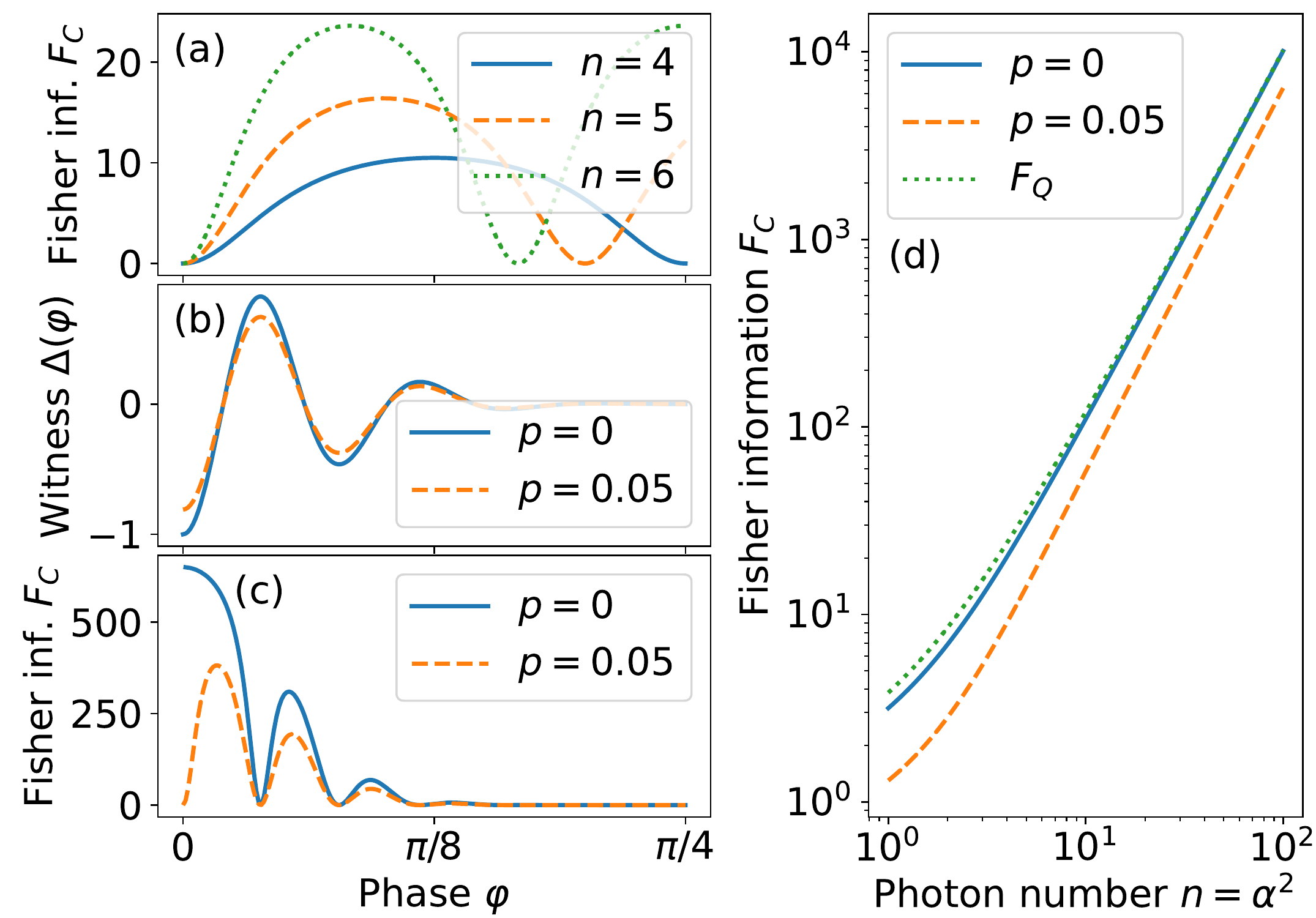}
	\caption{\label{fig:FisherFlip}
		Fisher information with ancilla phase-flip errors.
		(a) Classical Fisher information for NOON states with $n=4$ (solid blue line), $n=5$ (dashed orange line), and $n=6$ (dotted green line) with phase-flip probability of 5 \%.
		(b) Overlap witness for coherent states with $\alpha_1 = \alpha$, $\alpha_2 = 0$ with $\alpha = 5$ in the ideal case ($p=0$, solid blue line) and with phase flips (probability $p=0.05$, dashed orange line).
		(c) Classical Fisher information for the overlap witness shown in (b).
		(d) The maximum classical Fisher information (for coherent states with $\alpha_2 = 0$, optimized over the phase $\varphi$) as a function of the photon number $n=\alpha^2$;
		the quantum Fisher information is plotted as well (dotted green line).}
\end{figure}

Finally, we analyse the effect of phase-flip errors on the classical Fisher information in Fig.~\ref{fig:FisherFlip}.
As we describe in the Methods, the Fisher information with NOON states can be shown to be
\begin{equation}
	F_C^{\rm NOON}(\varphi) = \frac{(1-2p_1)^2(1-2p_2)^2n^2\sin^2(n\varphi)}{1-(1-2p_1)^2(1-2p_2)^2\cos^2(n\varphi)},
\end{equation}
which is plotted in panel (a) for $n=4,5,6$ and phase-flip probability $p = 0.05$.
Unlike the ideal case, the Fisher information is now phase-dependent with maximum reached for $\varphi = (2k+1)\pi/2n$ with $k\in\mathbb{Z}$, corresponding to the region where the overlap witness $\Delta(\varphi)$ can be approximated by a linear function of the phase.
This maximum is given by
\begin{equation}
	F_C^{\rm NOON} = (1-2p_1)^2(1-2p_2)^2n^2,
\end{equation}
preserving the Heisenberg scaling, albeit with a prefactor $(1-2p_1)^2(1-2p_2)^2$ that reduces the overall sensitivity.

For entangled coherent states (limiting ourselves only to the case of $\alpha_1 = \alpha$, $\alpha_2 = 0$), we first need to evaluate the probabilities in the second swap test (see Methods).
We plot the corresponding overlap witness in Fig.~\ref{fig:FisherFlip}(b), which shows that entangled coherent states suffer from a reduction of visibility that is quantitatively similar to NOON states.
The corresponding Fisher information is plotted in panel (c) for the ideal case (without phase flips, $p = 0$) and with phase-flip error probability of 5 \%.
Similar to the case of NOON states, the Fisher information becomes zero for $\varphi = 0$ when phase-flip errors are present.
The maximum Fisher information (optimized over the phase $\varphi$) is further investigated in Fig.~\ref{fig:FisherFlip}(d).
As expected, phase-flip errors reduce the Fisher information but this effect is rather small, especially for large photon numbers;
in this limit, the classical Fisher information is approximately equal to $(1-2p_1)^2(1-2p_2^2)n^2$, which is the same limit as for NOON states.

\subsection{Implementation in circuit QED}

The proposed swap-test interferometer can be implemented in circuit QED using the apparatus shown schematically in Fig.~\ref{fig:Imetry}(c).
It consists of two 3D microwave cavities (two microwave modes with annihilation operators $a,b$) both coupled to a superconducting nonlinear asymmetric inductive element (SNAIL)~\cite{Frattini2017}.
The SNAIL is a device exhibiting both third- and fourth-order nonlinearity which are both necessary for a CPBS gate with a cat-based ancilla.
Three-wave mixing (enabled by the third-order nonlinearity) is used for two-photon driving of the device which, together with the fourth-order Kerr nonlinearity, creates and stabilizes the cat qubit~\cite{Puri2017}.
Four-wave mixing (enabled by the Kerr nonlinearity) is then used to implement a cat-state-dependent beam splitter between the two fields which can be used to engineer a controlled-swap gate~\cite{Pietikainen2020}.

The ideal controlled-phase beam splitter between the two cavity fields controlled by the Kerr cat can be described by the effective Hamiltonian~\cite{Pietikainen2020}
\begin{equation}\label{eq:Heff}
	H_{\rm eff} = -Kc^{\dagger 2} c^2 + \epsilon c^{\dagger 2} + \epsilon^\ast c^2 + i\zeta_1(a^\dagger bc^\dagger - ab^\dagger c),
\end{equation}
where $c$ is the annihilation operator of the cat ancilla.
The first term describes the Kerr nonlinearity of the cat which, together with the two-photon driving (the second and third term), stabilizes the SNAIL in the subspace spanned by $\ket{\pm\beta}$, where $\beta = \sqrt{\epsilon/K}$~\cite{Grimm2020}.
Identifying the two coherent states as the logical qubit states (with $\ket{0_L} = \ket{+\beta}$, $\ket{1_L} = \ket{-\beta}$),
the last term in the Hamiltonian~\eqref{eq:Heff} describes (in a mean-field approximation where $\avg{c} = \pm\beta$) a beam-splitter interaction between the microwave cavity modes $a,b$ with a phase that depends on the logical state of the Kerr-cat qubit at a rate $\chi|\beta|$.
Importantly, the main decoherence mechanism for the Kerr cat is photon loss, which results predominantly in phase flips of the logical state, while bit flips are suppressed exponentially in the cat size $\beta^2$~\cite{Puri2019}.
Finally, measurement of the ancilla Kerr-cat qubit in the $X$ basis can be achieved by a series of qubit rotations followed by a conditional displacement of a readout resonator and homodyne detection~\cite{Puri2019,Grimm2020}. 
This measurement projects the ancilla onto one of the cat states $\ket{\pm_L} \propto \ket{\beta}\pm\ket{-\beta}$ which are the eigenstates of the logical $X$ operator~\cite{Sun2014,Ofek2016}.

\begin{figure}
	\centering
	\includegraphics[width=\linewidth]{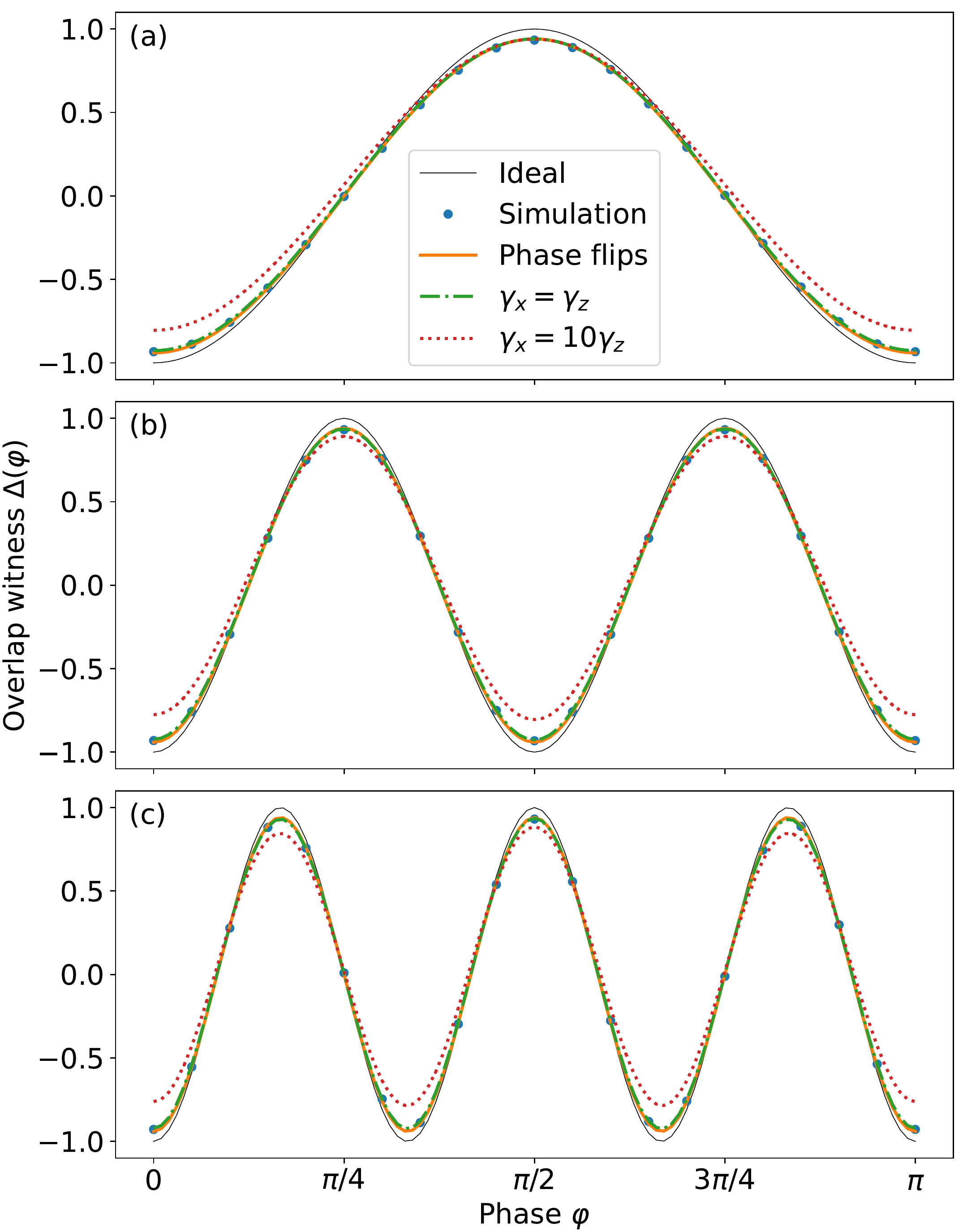}
	\caption{\label{fig:numerics}
		Numerical simulations of swap-test interferometry using NOON states with $n=2$ (a), $n=4$ (b), and $n=6$ (c).
		In all panels, we compare the results of numerical simulations (blue dots) with the ideal witness $\Delta(\varphi) = -\cos(n\varphi)$ (thin black line) and overlap witness with phase-flip errors included (thick orange line), $\Delta(\varphi) = -(1-2p)^2\cos(n\varphi)$, where $p = \kappa\beta^2\tau$ is the probability of an ancilla-phase flip error during the swap test.
		We also plot the overlap witness for a qubit model including bit flips at the same rate as phase flips ($\gamma_x = \gamma_z$, dot-dashed green line) and at a rate ten times higher ($\gamma_x = 10\gamma_z$, dotted red line).
		The interference visibility of the numerically simulated protocols is 93.3 \% ($n=2$), 93.2 \% ($n=4$), and 93.0 \% ($n=6$) which are all very close to the visibility with ancilla phase-flip errors only, $V = (1-2p)^2 = 94.0 \%$.}
\end{figure}

The beam-splitter operations corresponding to the two states of the cat ancilla are inverses of each other~\cite{Pietikainen2020}.
We can therefore combine a 50:50 controlled-phase beam splitter with an unconditional balanced beam splitter to engineer a controlled-beam splitter gate (see Methods):
for the logical state $\ket{0_L}$, the two operations exactly cancel each other, while for the ancilla in the state $\ket{1_L}$ they add up to a full swap of the two fields.
The only difference from an ideal controlled-swap gate is a conditional phase of $\pi$ that one of the fields acquires with each photon that is swapped.
For even-numbered NOON states, this phase is irrelevant as the total phase shift is always a multiple of $2\pi$;
for odd-numbered NOON states, the second swap test becomes insensitive to the relative phase between the two modes, resulting in probability $p_\pm = \frac{1}{2}$ independent of the input state.
Finally, for coherent states, this conditional phase shift modifies the overlap witness, leading to a reduced phase sensitivity compared to using a controlled-swap gate (see Methods);
crucially, this approach preserves the quadratic scaling of the classical Fisher information with the photon number, albeit with a smaller prefactor.
At the same time, this approach allows for a simplified experimental setup as the deterministic beam splitters need not be implemented as they transform coherent states into coherent states (Methods).

The overlap witness for the NOON states $|\Psi_-\rangle = (|n\rangle|0\rangle -|0\rangle|n\rangle)/\sqrt{2}$ with $n = 2,4,6$ is shown in Fig.~\ref{fig:numerics}.
For all three photon numbers, the results of the numerical simulation (blue dots; these include the effects of SNAIL decoherence and cross-Kerr interactions between the Kerr cat and the cavity modes as described in Methods) are very close to the simple phase-flip model (thick orange line), $\Delta(\varphi) = -(1-2p)^2\cos(n\varphi)$, where $p = \kappa\beta^2\tau \simeq 1.5 \%$ is the probability of an ancilla phase-flip error for single-photon-loss rate $\kappa$ and CPBS gate time $\tau$;
the observed interference visibility only weakly depends on the photon number (see figure caption for details).
This result implies that the classical Fisher information of this realistic device is also close to the Fisher information with phase flips with the maximum $F_C^{\rm NOON} = (1-2p)^4n^2$, guaranteeing Heisenberg scaling.
Additionally, we compare these results to a simple qubit model with bit flips to estimate the effect of this type of error.
As we describe in detail in Methods, this is achieved by replacing the (generally multilevel) Kerr cat with an ideal two-level system with phase-flip-error rate $\gamma_z = \kappa\beta^2$ and bit-flip-error rate $\gamma_x$.
In addition to the faster reduction of visibility with photon number, the interference fringes become asymmetric (the minimum increases faster than the maximum decreases), making fitting of experimental data more involved (as it requires fitting not only the visibility but also the offset of the centre of the interference fringe from zero).

\section{Discussion}

Apart from circuit QED, swap gates and swap tests have also been implemented with trapped ions~\cite{Gan2020,Nguyen2021ar} and so these two platforms provide ideal settings for swap-test interferometry.
In addition, circuit quantum acoustodynamics (QAD) uses the toolbox of circuit QED to control mechanical vibrations~\cite{Chu2020,Clerk2020} and can thus benefit from the same noise-biased gates as circuit QED platforms.
Mechanical degrees of freedom (available in circuit QAD and with trapped ions) readily interact with a broad range of physical systems and are therefore ideal for sensing weak forces and fields;
swap-test interferometry provides a new approach to detecting these forces with Heisenberg scaling.

In summary, we have presented a new approach to nonlinear interferometry based on swap tests.
Replacing linear beam splitters in a Mach--Zehnder interferometer by controlled-swap gates and measurement on ancilla qubits makes Heisenberg scaling attainable with simple input states---Fock and coherent states.
We presented a detailed analysis of ancilla qubit errors and established a crucial difference between phase- and bit-flip errors:
While the former reduce interference visibility and can, in principle, be corrected with repeated swap tests, the latter lead to imperfect swap operations, modifying the resulting interference pattern of the overlap witness and making the Heisenberg scaling unattainable.

This disparity between different types of qubit errors highlights the importance of qubits with biased noise.
These qubits recently attracted attention in the context of quantum computing where they offer a range of advantages in the design of quantum gates~\cite{Puri2020} and in quantum error correction~\cite{Puri2019,Darmawan2021}.
Building on these results, we proposed and analysed a possible implementation of swap-test interferometry with ancilla qubits based on Kerr cats which are strongly biased towards phase flips and thus fulfill the error requirements for approaching Heisenberg-limited phase sensitivity.
In this context, the proposed scheme can also be used to benchmark the performance of controlled-swap gates with ancilla qubits exhibiting biased noise~\cite{Pietikainen2020}.

Throughout the text, we considered only the effect of decoherence on the ancilla qubit and not on the fields.
This approach was motivated by two considerations:
First, the effect of losses on precision in linear Mach--Zehnder interferometry with NOON and entangled coherent states is well understood~\cite{DemkowiczDobrzanski2015,Haase2018};
in nonlinear swap-test interferometry, photon loss will play the same role during the free evolution of the fields between the swap tests.
Second, for state-of-the-art circuit QED systems, the lifetime of photons in three-dimensional microwave cavities can be one or two orders of magnitude longer than for on-chip nonlinear devices (such as the SNAIL used to implement the ancilla qubit)~\cite{Paik2011,Reagor2016,Chakram2021}, allowing evolution of the cavity fields with minimal decoherence in our proposed implementation.
For states containing $n$ photons, the loss rate is $\kappa_{\rm cav}n$, where $\kappa_{\rm cav}$ is the cavity dissipation rate;
for NOON states with tens of photons, the photon loss rate from the cavity modes can become comparable to the phase-flip rate of the Kerr-cat qubit.
With the required controlled-swap gates available in this platform, swap-test interferometry can thus be readily implemented with state-of-the-art experimental technology.

\section{Methods}

\subsection{Ancilla phase flips}

To account for phase flips of the ancilla in swap-test interferometry with NOON states, we first note that the controlled-swap gate is transparent to this type of error.
(Note that the Kraus operators describing the controlled-swap gate, $\frac{1}{2}(I-Z)\otimes S$, and a phase-flip error, $Z$, commute.)
We can therefore consider only the effect of phase flips just before the measurement.
A phase flip (with probability $p_1\ll 1$) during state preparation then results in incorrectly assigning the opposite meaning to the measurement result;
for the outcome $\ket{+}$, the antisymmetric singlet state $\ket{\Psi_-}=(\ket{n}\ket{0}-\ket{0}\ket{n})/\sqrt{2}$ is prepared while the symmetric state $\ket{\Psi_+}=(\ket{n}\ket{0}+\ket{0}\ket{n})/\sqrt{2}$ is prepared for the outcome $\ket{-}$.
Generally, the first swap test and postselection on the $\ket{-}$ state of the ancilla gives the mixed state
\begin{equation}\label{eq:PhaseFlipFock}
	\rho = (1-p_1)\ket{\Psi_-}\bra{\Psi_-} + p_1\ket{\Psi_+}\bra{\Psi_+}.
\end{equation}

After the phase shift $\varphi$, an ideal second swap test projects the fields onto the symmetric or antisymmetric subspace with probability (the calculation is straightforward but the expressions for the resulting states cumbersome so we do not include them here)
\begin{equation}\label{eq:Mprobs}
	p_\pm = \frac{1}{2}\pm\frac{2p_1-1}{2}\cos(n\varphi).
\end{equation}
The witness we obtain from these probabilities is given by
\begin{equation}
	\Delta(\varphi) = -(1-2p_1)\cos(n\varphi).
\end{equation}
For a phase flip with probability $p_2\ll 1$ during the second swap test,
the probabilities are modified according to
\begin{equation}\label{eq:Mprobs2}
	p_+\to(1-p_2)p_+ + p_2p_-,\quad p_-\to (1-p_2)p_- + p_2p_+;
\end{equation}
with probability $1-p_2$, no phase flip took place and the probabilities are unaffected, while a phase flip occurred with probability $p_2$ and the probabilities are flipped as well.
The total witness thus becomes
\begin{equation}
	\Delta(\varphi) = -(1-2p_1)(1-2p_2)\cos(n\varphi).
\end{equation}
The phase flips therefore reduce the visibility of the interference fringes.

\subsection{Swap-test interferometry with general pure states}

With general, overlapping pure states ($\sprod{\psi}{\phi} = s \in\mathbb{C}$), the first swap test again conditionally prepares one of the Bell-like states
\begin{equation}
	\ket{\Psi_\pm}\propto (\ket{\psi}\ket{\phi}\pm\ket{\phi}\ket{\psi}).
\end{equation}
The normalization and probability of generating these states can be found from the scalar product;
the normalization factor is
\begin{equation}
	N_\pm = \sprod{\Psi_\pm}{\Psi_\pm} = 2(1\pm|s|^2)
\end{equation}
and the probability $p_\pm = \frac{1}{2}(1\pm|s|^2)$.
Next, the state (we work again with $\ket{\Psi_-}$) acquires a phase shift $\varphi$ on the first mode,
\begin{equation}
	\ket{\Psi_-(\varphi)} = \frac{1}{\sqrt{N_-}}(\ket{\psi(\varphi)}\ket{\phi}-\ket{\phi(\varphi)}\ket{\psi}),
\end{equation}
where we denote the general phase-shifted state as $\ket{\chi(\varphi)} = e^{-i\varphi a^\dagger a}\ket{\chi}$ (with $\chi = \psi,\phi$) and $a$ is the annihilation operator of the first mode.

Afterwards, we apply the second swap test to estimate the overlap witness $\Delta(\varphi)$.
The controlled-swap gate transforms the state $\ket{\Psi_-(\varphi)}$ into
\begin{equation}
	\ket{\Psi_{\rm out}} = \frac{1}{2}\sqrt{\frac{M_+}{N_-}}\ket{+}\ket{\Omega_+(\varphi)} + \frac{1}{2}\sqrt{\frac{M_-}{N_-}}\ket{-}\ket{\Omega_-(\varphi)},
\end{equation}
where we introduced the field states
\begin{align}
\begin{split}
	\ket{\Omega_\pm(\varphi)} &= \frac{1}{\sqrt{M_\pm}}[\ket{\psi(\varphi)}\ket{\phi} - \ket{\phi(\varphi)}\ket{\psi}\\
	&\qquad\qquad \pm \ket{\phi}\ket{\psi(\varphi)}\mp \ket{\psi}\ket{\phi(\varphi)}]
\end{split}
\end{align}
with the normalization constant
\begin{align}\label{eq:MM}
\begin{split}
	M_\pm &= 4(1-|s|^2) \pm 2[|s(\varphi)|^2+|s(-\varphi)|^2]\\
	&\quad \mp 2[s_\psi(\varphi)s_\phi(-\varphi) + s_\psi(-\varphi)s_\phi(\varphi)].
\end{split}
\end{align}
The parameters in this expressions are defined via the scalar products
\begin{align}\label{eq:Mproducts}
\begin{split}
	s(\varphi) &= \sprod{\phi}{\psi(\varphi)} = \expct{\phi}{e^{-i\varphi a^\dagger a}}{\psi}, \\
	s_\chi(\varphi) &= \sprod{\chi}{\chi(\varphi)} = \expct{\chi}{e^{-i\varphi a^\dagger a}}{\chi}.
\end{split}
\end{align}
We can now obtain the overlap witness as the difference of the probabilities of finding the ancilla in the $\ket{+}$ and $\ket{-}$ states,
\begin{equation}\label{eq:Mwitness}
	\Delta(\varphi) = p_+ - p_-  = \frac{M_+-M_-}{4N_-}
\end{equation}
which gives Eq.~\eqref{eq:witness}.
Note that by virtue of the definitions~\eqref{eq:Mproducts}, we have $s_\chi^\ast(\varphi) = s_\chi(-\varphi)$ and the overlap witness~\eqref{eq:Mwitness} is always real as expected.

\subsection{Quantum and classical Fisher information}

For a pure quantum state $\ket{\psi}$, the quantum Fisher information can be calculated as~\cite{Liu2019}
\begin{equation}
	F_Q = 4\left(\left\langle\frac{\partial}{\partial\varphi}\psi\right.\left|\frac{\partial}{\partial\varphi}\psi\right\rangle - \left|\left\langle\frac{\partial}{\partial\varphi}\psi\bigg|\psi\right\rangle\right|^2\right).
\end{equation}
For NOON states, we have
\begin{equation}
	\left|\frac{\partial}{\partial\varphi}\psi\right\rangle = -\frac{in}{\sqrt{2}} e^{-in\varphi}\ket{n}\ket{0};
\end{equation}
a straightforward calculation then gives $F_Q^{\rm NOON} = n^2$.
For entangled coherent states, we express the coherent states in the Fock basis to obtain
\begin{equation}
	\left|\frac{\partial}{\partial\varphi}\alpha e^{-i\varphi}\right\rangle = -i\exp\left(-\frac{|\alpha|^2}{2}\right)\sum_{n=0}^\infty\frac{n\alpha^n e^{-in\varphi}}{\sqrt{n!}}\ket{n}.
\end{equation}
With this expression, we can evaluate scalar products of the form $\sprod{\partial\alpha e^{-i\varphi}/\partial\varphi}{\beta}$ and $\sprod{\partial\alpha e^{-i\varphi}/\partial\varphi}{\partial\beta e^{-i\varphi}/\partial\varphi}$ and get the general expression for the quantum Fisher information with general real amplitudes $\alpha_{1,2}\in\mathbb{R}$,
\begin{widetext}
\begin{equation}
	F_Q^{\alpha_1,\alpha_2} = 2\frac{\alpha_1^2+\alpha_1^4+\alpha_2^2+\alpha_2^4-2e^{-(\alpha_1-\alpha_2)^2}\alpha_1\alpha_2(1+\alpha_1\alpha_2)}{1-e^{-(\alpha_1-\alpha_2)^2}}
	-\frac{(\alpha_1^2+\alpha_2^2-2e^{-(\alpha_1-\alpha_2)^2}\alpha_1\alpha_2)^2}{\left(1-e^{-(\alpha_1-\alpha_2)^2}\right)^2}.
\end{equation}
For the two choices $\alpha_1 = \alpha = -\alpha_2$ and $\alpha_1 = \alpha$, $\alpha_2=0$, this expression simplifies to Eqs.~\eqref{eq:QFIecs}.

The classical Fisher information can be found from the probability of detecting the ancilla qubit in the second swap test in the state $\ket{\pm}$ using~\cite{Polino2020}
\begin{equation}
	F_C = p_+\left(\frac{\partial}{\partial\varphi}\ln p_+\right)^2 + p_-\left(\frac{\partial}{\partial\varphi}\ln p_-\right)^2,
\end{equation}
where both probabilities $p_\pm$ implicitly depend on the phase $\varphi$.
For NOON states, the probabilities are $p_\pm = \frac{1}{2}[1\pm\cos(n\varphi)]$, which give the classical Fisher information
\begin{equation}
	F_C^{\rm NOON} = n^2 = F_Q^{\rm NOON}.
\end{equation}
For coherent states, the probabilities  are given by $p_\pm = M_\pm/(4N_-)$, where $M_\pm$ are in Eq.~\eqref{eq:MM}.
The classical Fisher information is, unlike the quantum Fisher information,  phase dependent but the general expression is too complicated to be reproduced here;
for the two cases discussed above, we obtain
\begin{subequations}
\begin{align}
	F_C^{\alpha,-\alpha}(\varphi) &= \frac{4e^{4\alpha^2}[1+\exp(4\alpha^2\cos\varphi)]^2\alpha^4\sin^2\varphi}{-e^{4\alpha^2}+\exp(4\alpha^2\cos\varphi)+\exp[4\alpha^2(2+\cos\varphi)]-\exp[4\alpha^2(1+2\cos\varphi)]},\\
	F_C^{\alpha,0}(\varphi) &= \frac{\exp(2\alpha^2\cos\varphi)\alpha^4\sin^2(\varphi+\alpha^2\sin\varphi)}{[e^{\alpha^2}-\exp(\alpha^2\cos\varphi)\cos(\alpha^2\sin\varphi)][-2+e^{\alpha^2}+\exp(\alpha^2\cos\varphi)\cos(\alpha^2\sin\varphi)]}.
\end{align}
\end{subequations}

\subsection{Fisher information with phase-flip errors}

To analyse the effect of phase-flip errors on the estimation sensitivity, we evaluate the classical Fisher information in the presence of phase-flip errors.
Using Eqs.~\eqref{eq:Mprobs}, \eqref{eq:Mprobs2}, we can directly evaluate the classical Fisher information for NOON states,
\begin{equation}
	F_C^{\rm NOON}(\varphi) = \frac{(1-2p_1)^2(1-2p_2)^2n^2\sin^2(n\varphi)}{1-(1-2p_1)^2(1-2p_2)^2\cos^2(n\varphi)}.
\end{equation}
It is then straightforward to show that the minimum is reached for $\varphi_{\rm min} = k\pi/n$, where $k\in\mathbb{Z}$; we then have $F_C^{\rm NOON}(\varphi_{\rm min})= 0$. 
The maximum is achieved for $\varphi_{\rm max} = (2k+1)\pi/2n$ with $k\in\mathbb{Z}$ and is given by
\begin{equation}
	F_C^{\rm NOON}(\varphi_{\rm max}) = (1-2p_1)^2(1-2p_2)^2n^2,
\end{equation}
preserving the Heisenberg scaling, albeit with a prefactor $(1-2p_1)^2(1-2p_2)^2$ that reduces the overall sensitivity.

For phase estimation with entangled coherent states, we first evaluate the probabilities in the second swap test.
Following the same procedure as for NOON states (for which we obtained Eqs.~\eqref{eq:Mprobs}, \eqref{eq:Mprobs2}), we get the probability of the measurement outcome $\pm$ for two general pure states with overlap $s = \sprod{\psi}{\phi}$,
\begin{align}
	p_\pm &= \left(\frac{1-p_1}{N_-}+\frac{p_1}{N_+}\right)\left(1\pm\frac{|s(\varphi)|^2+|s(-\varphi)|^2}{2}\right) 
		+\left(\frac{p_1}{N_+}-\frac{1-p_1}{N_-}\right)\left(|s|^2 \pm \frac{s_\psi(\varphi)s_\phi(-\varphi)+s_\psi(-\varphi)s_\phi(\varphi)}{2}\right). 
\end{align}\pagebreak
\end{widetext}

\noindent
Focusing on the case $\alpha_1 = \alpha\in\mathbb{R}$, $\alpha_2 = 0$, a straightforward calculation gives
\begin{align}\label{eq:WitnessFlips}
\begin{split}
	\Delta(\varphi) &= \frac{1-\exp(\alpha^2\cos\varphi)\cos(\alpha^2\sin\varphi)}{e^{\alpha^2}-1} \\
		&\quad -2p_1\frac{1-\exp[\alpha^2(1+\cos\varphi)]\cos(\alpha^2\sin\varphi)}{e^{2\alpha^2}-1}.
\end{split}
\end{align}
Phase-flip errors during the first swap test thus give rise to a more general modification of the interference pattern than in the case of NOON states where it gives a constant factor $1-2p_1$.
Phase flips during the second swap test, on the other hand, act the same way as before, resulting in a constant factor $1-2p_2$ multiplying the overlap witness, $\Delta(\varphi)\to(1-2p_2)\Delta(\varphi)$.
From these probabilities, one can also obtain an analytical expression for the classical Fisher information;
we do not reproduce it here as it is long and provides no insight.

\subsection{Implementation in circuit QED}

In the mean-field approximation (where we replace the operators for the cat qubit with their classical value, $\avg{c} = \avg{c^\dagger} = \pm\beta\in\mathbb{R}$), the ideal controlled-phase beam splitter Hamiltonian~\eqref{eq:Heff} becomes
\begin{equation}
	H_\pm = \pm i\chi\beta(a^\dagger b - b^\dagger a).
\end{equation}
This Hamiltonian describes beam-splitter coupling between the two cavity modes at a rate $\chi\beta$.
These transformations are described by the unitaries
\begin{subequations}
\begin{align}
	U_+ &= U_-^\dagger = \begin{pmatrix}
		t & r \\ -r & t
	\end{pmatrix}, \\
	t &= \cos(\chi\beta\tau), \\
	r &= \sin(\chi\beta\tau),
\end{align}
\end{subequations}
where $\tau$ is the duration of the interaction.
The transformation of the fields is described by $(a_,b)^T \to U_\pm (a,b)^T$.

\begin{figure}
\centering
\includegraphics[width=\linewidth]{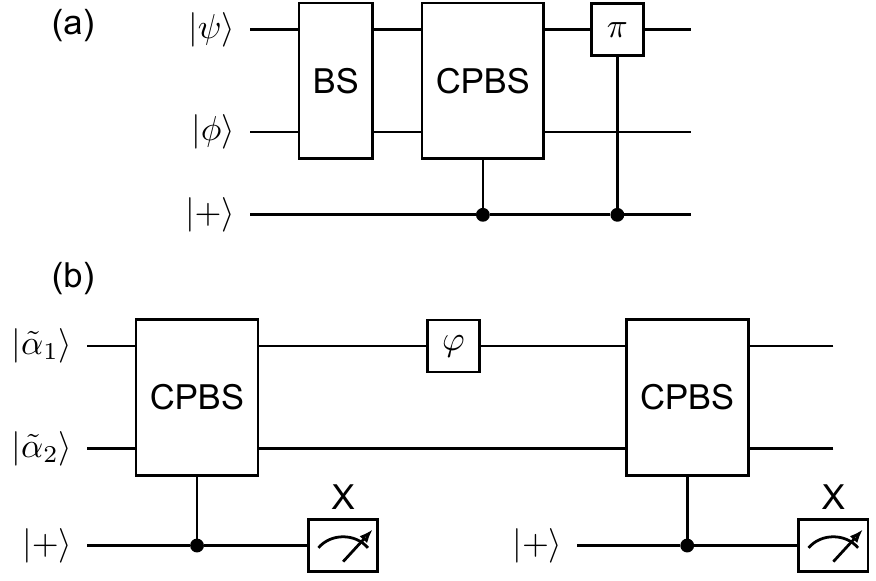}
\caption{\label{fig:cSWAP} Swap-test interferometry with controlled-phase beam splitter gates. (a) Circuit for implementing a controlled-swap gate using a 50:50 beam splitter gate followed by a 50:50 controlled-phase beam splitter and a controlled-phase gate.
	Alternatively, the order of the beam splitter and controlled-phase beam splitter can be exchanged.
	(b) Scheme for swap-test interferometry with balanced controlled-phase beam splitter gates and ancilla Kerr-cat qubits in circuit QED.
	Deterministic beam splitters are not needed when starting from a suitably modified initial coherent states with $\tilde{\alpha}_{1,2} = (\alpha_1\mp\alpha_2)/\sqrt{2}$.}
\end{figure}

The CPBS interaction can be used to implement a controlled-swap gate using the circuit in Fig.~\ref{fig:cSWAP}.
A balanced CPBS gate (i.e., a gate with $t = r = 1/\sqrt{2}$) is preceded (or, equivalently, followed) by a deterministic beam splitter that applies the unitary $U_-$.
When the cat ancilla is in the logical state $\ket{0_L} = \ket{\beta}$, the two gates cancel each other since $U_+ = U_-^\dagger$ and the joint state of the fields is unchanged.
When, on the other hand, the cat starts from the logical state $\ket{1_L} = \ket{-\beta}$, the two gates add up and perform a full swap of the two fields.
The final controlled-phase gate (a $\pi$ shift of the first mode) compensates the relative phase that the field acquires during the beam-splitter transformation.
The circuit thus implements the unitary $U_{\rm cswap} = \ket{0_L}\bra{0_L}\otimes I + \ket{1_L}\bra{1_L}\otimes S$, where $I$ is the identity and
\begin{equation}
	S = \begin{pmatrix}
		0 & 1 \\ 1 & 0
	\end{pmatrix}
\end{equation}
is the swap unitary.

\begin{figure}
	\centering
	\includegraphics[width=\linewidth]{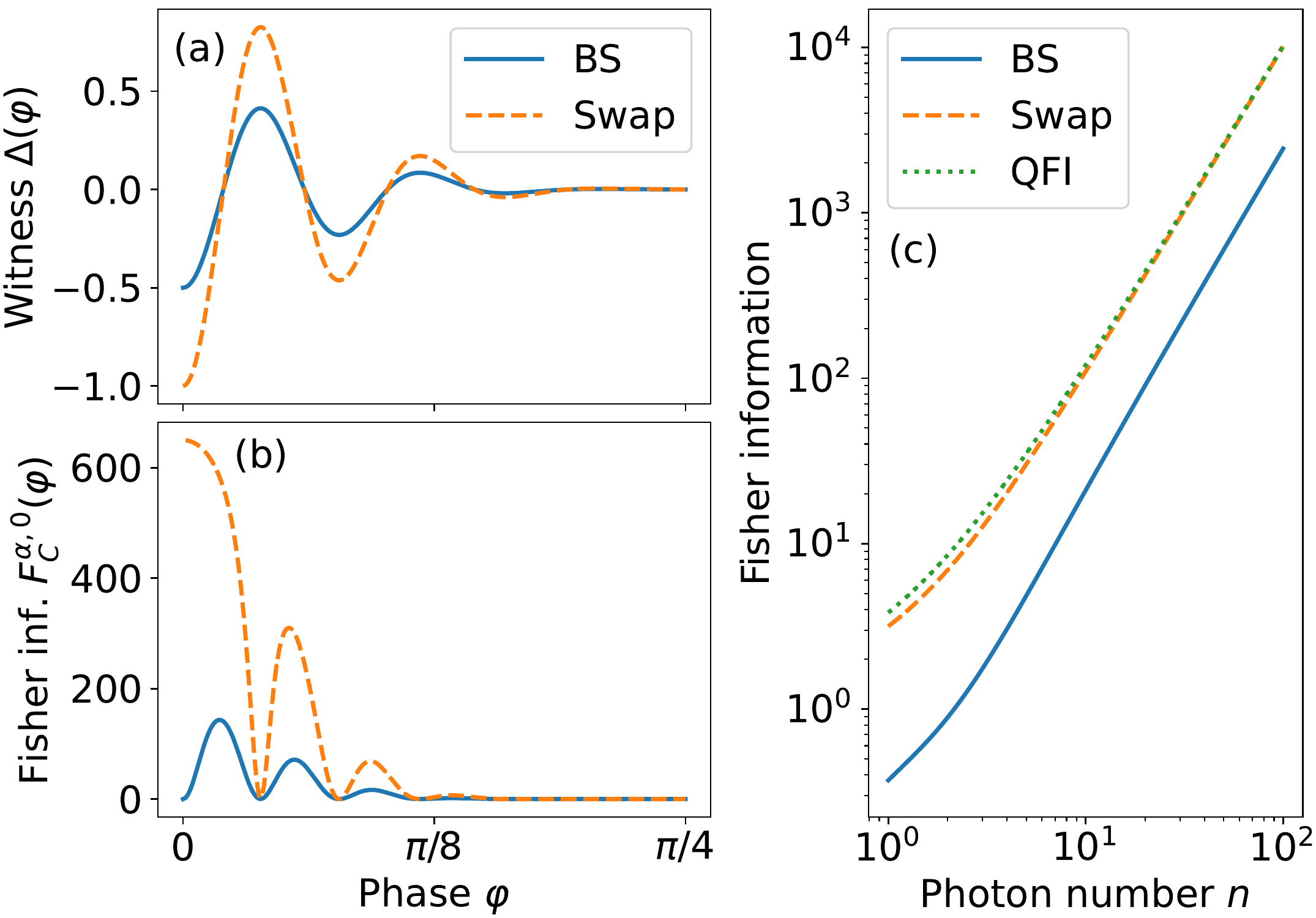}
	\caption{\label{fig:CBS}Swap-test interferometry with controlled beam splitters.
	(a) Overlap witness $\Delta(\varphi)$ for controlled beam splitter (BS, solid blue line) and controlled swap (dashed orange line) for coherent state with $\alpha = 5$ and the vacuum.
	(b) Classical Fisher information $F_C^{\alpha,0}$ corresponding to the curves in panel (a).
	(c) Maximum of the classical Fisher information over phase plotted against the photon number $n = \alpha^2$.
	The dotted green line shows the quantum Fisher information.}
\end{figure}

In an experiment, the conditional phase gate can be omitted with little to no penalty in terms of sensitivity.
For an input with a specific photon number $n$, the phase associated with the beam splitter gives a total phase $(-1)^n$, which is irrelevant for states with an even photon number.
For coherent states (we consider the case $\alpha_1 = \alpha\in\mathbb{R}$, $\alpha_2 = 0$ since it is the optimal scenario), a straightforward calculation reveals the modified overlap witness
\begin{equation}
	\Delta(\varphi) = -\frac{1 - \cosh(\alpha^2\cos\varphi)\cos(\alpha^2\sin\varphi)}{1-\exp(\alpha^2)}
\end{equation}
which we compare with the ideal case of Eq.~\eqref{eq:Delta_a0} in Fig.~\ref{fig:CBS}(a).
The overlap witness now oscillates only between $\pm\frac{1}{2}$ due to the negligible overlap between the states $\ket{\pm\alpha}$ for large $\alpha$;
this additional phase shift also leads to the oscillations reappearing around $\varphi = \pi$ (not shown in the plot).
Despite this modified behaviour, the quantum Fisher information $F_Q^{\alpha,0}$ remains unchanged when the controlled-swap gate is replaced by a controlled beam splitter, allowing, in principle, the same Heisenberg-limited phase sensitivity as with controlled-swap gates.
The classical Fisher information,
\begin{widetext}
\begin{equation}
	F_C^{\alpha,0}(\varphi) = \frac{\alpha^4[\cosh(\alpha^2\cos\varphi)\sin(\alpha^2\sin\varphi)\cos\varphi + \sinh(\alpha^2\cos\varphi)\cos(\alpha^2\sin\varphi)\sin\varphi]^2}{[e^{\alpha^2}-\cosh(\alpha^2\cos\varphi)\cos(\alpha^2\sin\varphi)][-2+e^{\alpha^2}+\cosh(\alpha^2\cos\varphi)\cos(\alpha^2\sin\varphi)]},
\end{equation}	
\end{widetext}
is, however, reduced as can be seen in Fig.~\ref{fig:CBS}(b) which shows that the maximum Fisher information shifts from $\varphi = 0$ we had with controlled-swap gates.
This maximum is reduced but still keeps the quadratic scaling in the photon number as shown in Fig.~\ref{fig:CBS}(c).

For experimental implementation with coherent states, a further simplification is possible by omitting the deterministic beam-splitter gates.
Since coherent states are transformed by linear beam splitters onto coherent states, one can start with modified input states,
\begin{equation}
	\begin{pmatrix}
		\tilde{\alpha}_1 \\ \tilde{\alpha}_2
	\end{pmatrix} = U_-
	\begin{pmatrix}
		\alpha_1 \\ \alpha_2 
	\end{pmatrix} = \frac{1}{\sqrt{2}}
	\begin{pmatrix}
		\alpha_1-\alpha_2 \\ \alpha_1+\alpha_2 
	\end{pmatrix},
\end{equation}
instead of applying the first deterministic beam splitter on the input states $\alpha_{1,2}$.
The deterministic beam splitter of the second swap test can be applied after the controlled-phase beam splitter;
since we are interested only in the statistics of the ancilla measurement (which are unaffected by the beam splitter), the deterministic beam splitter is irrelevant;
with coherent states, the simplified interferometer that is shown in Fig.~\ref{fig:cSWAP}(b) can therefore be used.

\subsection{Numerical simulations}

We simulate the swap-test interferometer using the controlled beam splitter (CBS) introduced in Ref.~\cite{Pietikainen2020}. The CBS operation is described by the Hamiltonian
\begin{subequations}
\begin{align} \label{eq:Ham}
	{H} &= {H}_{0} + {H}_{\rm CPBS} + {H}_{\rm BS}, \\
	\begin{split}
		{H}_0 &= -K{c}^{\dagger 2}{c}^2 +\epsilon{c}^{\dagger 2} +\epsilon^\ast {c}^{2}\\
			&\quad -\chi ({a}^\dagger{a} + {b}^\dagger{b}-N)({c}^\dagger {c}-|\alpha|^2),
	\end{split} \\
	{H}_{\rm CPBS} &= -\zeta_1{a}^\dagger{b}{c}^\dagger -\zeta_1^\ast{a}{b}^\dagger{c}, \\
	{H}_{\rm BS} &= \zeta_2{a}^\dagger{b} + \zeta_2^\ast {a}{b}^\dagger,
\end{align}
\end{subequations}
where ${H}_0$ describes the Kerr-cat ancilla and the cross-Kerr interactions between the Kerr cat and the microwave modes, ${H}_{\rm CPBS}$ is the controlled-phase beam splitter (CPBS) coupling and ${H}_{\rm BS}$ is the deterministic beam splitter (BS) coupling.
The CPBS and BS interactions are switched on and off sequentially (the CPBS is applied first) and the duration of each interaction is chosen to give rise to a balanced (50:50) splitting ratio.
To decrease ancilla errors during the swap tests, we perform the ancilla measurements after the CPBS and before the BS interaction.

The full dynamics during a swap test are described by the master equation
\begin{equation}\label{eq:ME}
	\dot{{\rho}} = -i[{H},{\rho}] +\kappa(1+N_t)\mathcal{D}[{c}]{\rho} +\kappa N_t\mathcal{D}[{c}^\dagger]{\rho} +\kappa_2\mathcal{D}[{c}^2]{\rho} \,,
\end{equation}
where $\mathcal{D}[{o}]{\rho} = {o}{\rho}{o}^\dagger -\frac{1}{2}{o}^\dagger{o}{\rho} -\frac{1}{2}{\rho}{o}^\dagger{o}$ is the Lindblad superoperator, $\kappa$ and $\kappa_2$ are the single- and two-photon dissipation rates of the ancilla, and $N_t$ is the thermal population of the Kerr-cat mode.
The two-photon dissipation term $\kappa_2\mathcal{D}[c^2]\rho$ is added to help stabilize the Kerr cat within the qubit subspace~\cite{Pietikainen2020}.
The parameters for simulations are similar to the recent experimental demonstration of a stabilized Kerr-cat qubit~\cite{Grimm2020,Pietikainen2020} and are summarized in Table~\ref{tab:params}.

\begin{table}
\centering
\caption{\label{tab:params}System parameters for numerical simulations}
\begin{tabular}{|p{0.4\linewidth}p{0.3\linewidth}p{0.2\linewidth}|}
	\hline
	Parameter & Symbol & Value \\
	\hline
	Kerr nonlinearity & $K/2\pi$ & \SI{6.7}{\mega\hertz} \\
	Two-photon driving & $\epsilon/2\pi$ & \SI{20.1}{\mega\hertz} \\
	Kerr-cat amplitude & $\beta = \sqrt{\epsilon/K}$ & $\sqrt{3}$ \\
	Cross-Kerr coupling & $\chi/2\pi$ & \SI{603}{\kilo\hertz} \\
	CPBS rate & $\zeta_1/2\pi$ & \SI{120}{\kilo\hertz} \\
	BS rate & $\zeta_2/2\pi=\zeta_1\beta/2\pi $ & \SI{210}{\kilo\hertz} \\
	Single-photon loss & $\kappa/2\pi$ & \SI{1.35}{\kilo\hertz} \\
	Thermal occupation & $N_t$ & 0.06 \\
	Two-photon loss & $\kappa_2/2\pi$ & \SI{135}{\kilo\hertz} \\
	CPBS gate time & $\tau$ & \SI{600}{\nano\second} \\
	Phase-flip probability & $p = \kappa\beta^2\tau$ & 1.5 \% \\
	\hline
\end{tabular}
\end{table}

To evaluate the effect of bit-flip errors, we have devised a toy model where the Kerr cat ancilla is replaced with a two-level system.
For this qubit model, we use the effective CPBS Hamiltonian
\begin{equation}
{H}_{\rm CPBS} = -\zeta_1\beta({a}^\dagger{b} -{a}{b}^\dagger)Z,
\end{equation}
with the master equation
\begin{equation}
\dot{{\rho}} = -i[{H}_{\rm CPBS}+H_{\rm BS},{\rho}] +\gamma_z\mathcal{D}[Z]{\rho} +\gamma_x\mathcal{D}[X]{\rho},
\end{equation}
where the deterministic BS Hamiltonian is the same as before and $\gamma_{z,x}$ are the phase- and bit-flip error rate;
we set $\gamma_z = \kappa\beta^2$ to have the same phase-flip error rate as for the full simulation with the Kerr-cat ancilla.

%

\section{Acknowledgments}
	O.\v{C}. and I.P. have received funding from the project LTAUSA19099 of the Czech Ministry of Education, Youth and Sports (MEYS \v{C}R).
	R.F. acknowledges project 21-13265X of the Czech Science Foundation.
	O.\v{C}., I.P., and R.F. have been further supported by the European Union’s 2020 research and innovation programme (CSA - Coordination and support action, H2020-WIDESPREAD-2020-5) under grant agreement No. 951737 (NONGAUSS).
	S.P. and S.M.G. acknowledge support by the 
	Army Research Office under award W911NF-18-1-0212.
	The views and conclusions contained in this document are those of the authors and should not be interpreted as representing the official policies, either expressed or implied, of the Army Research Office (ARO), or the U.S. Government. 
	The U.S. Government is authorized to reproduce and distribute reprints for Government purposes notwithstanding any copyright notation herein.

\section{Author contributions}
O.\v{C}. and R.F. developed the idea and performed theoretical analyses.
O.\v{C}. and I.P. designed and performed numerical simulations.
O.\v{C}. wrote the manuscript with input from all authors.

\section{Competing interests}
The authors declare no competing interests.

\end{document}